\documentclass[journal]{IEEEtran}

\usepackage{graphicx}
\usepackage{subcaption}
\usepackage{cite}
\usepackage{epstopdf}
\usepackage{bm}
\usepackage{multirow}
\usepackage[switch]{lineno}
\usepackage[colorlinks=true, allcolors=blue]{hyperref}
\usepackage{epstopdf}
\usepackage{color,soul}
\usepackage{subcaption}
\usepackage{mathtools}
\usepackage{soul}
\DeclarePairedDelimiter\ceil{\lceil}{\rceil}
\DeclarePairedDelimiter\floor{\lfloor}{\rfloor}

\begin{document}
\title{Predicting Network Controllability Robustness:\\
       {\Huge A Convolutional Neural Network Approach} }

\author{Yang~Lou,~\IEEEmembership{}
        Yaodong~He,~\IEEEmembership{}
        Lin~Wang,~\IEEEmembership{Senior~Member,~IEEE,}
        and~Guanrong~Chen,~\IEEEmembership{Life~Fellow,~IEEE}%
\thanks{Y. Lou, Y. He, and G. Chen are with the Department of Electrical Engineering, City University of Hong Kong (e-mails: felix.lou@my.cityu.edu.hk; yaodonghe2-c@my.cityu.edu.hk; eegchen@cityu.edu.hk).}%
\thanks{L. Wang is with the Department of Automation, Shanghai Jiao Tong University, Shanghai 200240, China, and also with the Key Laboratory of System Control and Information Processing, Ministry of Education, Shanghai 200240, China (e-mail: wanglin@sjtu.edu.cn).}%
\thanks{This research was supported in part by the National Natural Science Foundation of China under Grant 61873167, and in part by the Natural Science Foundation of Shanghai under Grand 17ZR1445200.}%
\thanks{(\textit{Corresponding author: Guanrong Chen.})}
\thanks{\color{blue}This paper has been published in \textit{IEEE Transactions on Cybernetics}.}
\thanks{\url{https://doi.org/10.1109/TCYB.2020.3013251}}
}

\markboth{\url{https://doi.org/10.1109/TCYB.2020.3013251} (September~2020)}%
{Lou \MakeLowercase{\textit{et al.}}: Predicting Network Controllability Robustness}

\maketitle
\begin{abstract}
Network controllability measures how well a networked system can be controlled to a target state, and its robustness reflects how well the system can maintain the controllability against malicious attacks by means of node-removals or edge-removals. The measure of network controllability is quantified by the number of external control inputs needed to recover or to retain the controllability after the occurrence of an unexpected attack. The measure of the network controllability robustness, on the other hand, is quantified by a sequence of values that record the remaining controllability of the network after a sequence of attacks. Traditionally, the controllability robustness is determined by attack simulations, which is computationally time consuming. In this paper, a method to predict the controllability robustness based on machine learning using a convolutional neural network is proposed, motivated by the observations that 1) there is no clear correlation between the topological features and the controllability robustness of a general network, 2) the adjacency matrix of a network can be regarded as a gray-scale image, and 3) the convolutional neural network technique has proved successful in image processing without human intervention. Under the new framework, a fairly large number of training data generated by simulations are used to train a convolutional neural network for predicting the controllability robustness according to the input network-adjacency matrices, without performing conventional attack simulations. Extensive experimental studies were carried out, which demonstrate that the proposed framework for predicting controllability robustness of different network configurations is accurate and reliable with very low overheads.
\end{abstract}


\IEEEpeerreviewmaketitle

\section{Introduction}
\label{sec:intro}

\IEEEPARstart{C}{omplex} networks have gained wide popularity and rapid development during the last two decades. Scientific research on this subject was pursued with great efforts from various scientific and engineering communities, which has literally become a self-contained discipline interconnecting network science, systems engineering, statistical physics, applied mathematics and social sciences alike \cite{Barabasi2016NS,Newman2010N,Chen2014Book}.

Recently, the network controllability issue has become a focal topic in complex network studies \cite{Liu2011N,Yuan2013NC,Posfai2013SR,Menichetti2014PRL,Motter15CHAOS,Wang2016AUTO,Liu2016RMP,Wang2017RSPTA,Wang20L17SR,Zhang2017TAC}, where the concept of \textit{controllability} refers to the ability of a network in moving from any initial state to any target state under an admissible control input within a finite duration of time. It was shown that identifying the minimum number of external control inputs (recalled driver nodes) to achieve a full control of a directed network amounts to searching for the maximum matching of the network, known as the \textit{structural controllability} \cite{Liu2011N}. Along the same line of research, in \cite{Yuan2013NC}, an efficient tool to assess the \textit{state controllability} of a large-scale network is suggested. In \cite{Menichetti2014PRL}, it reveals that random networks are controllable by an infinitesimal fraction of driver nodes, if both of its minimum in-degree and out-degree are greater than two. In \cite{Posfai2013SR}, it demonstrates that clustering and modularity have no discernible impact on the network controllability, while degree correlations have certain effects. The network controllability of some canonical graph models is studied quite thoroughly in \cite{Wu2018JNS}. Moreover, the controllability of multi-input/multi-output networked systems is studied in \cite{Wang2016AUTO,Hao2018IJRNC}. A comprehensive overview of the subject is available in the recent survey papers \cite{Xiang2019CSM,Wen2020TSMC}.

Meanwhile, malicious attacks on complex networks is a main concerned issue today \cite{Holme2002PRE,Shargel2003PRL,Schneider2011PNAS,Liu2012PO,BBBH13,Xiao2014CPB}. Reportedly, degree-based node attacks, by means of removing nodes with high degrees, are more destructive than random attacks on network controllability over directed random-graph and scale-free networks \cite{Pu2012PA}. The underlying hierarchical structure of such a network leads to the effective random upstream (or downstream) attack, which removes the hierarchical upstream (or downstream) node of a randomly picked one, since this attack removes more hubs than a random attack \cite{Liu2012PO}. Both random and intentional edge-removal attacks have also been studied by many. In \cite{Nie2014PO}, for example, it shows that the intentional edge-removal attack by removing highly-loaded edges is very effective in reducing the network controllability. It is moreover observed (e.g., in \cite{BPPSH10}) that intentional edge-based attacks are usually able to trigger cascading failures in scale-free networks but not necessarily in random-graph networks. These observations have motivated some recent in-depth studies of the robustness of network controllability \cite{Lou2018TCASI}. In this regard, both random and intentional attacks as well as both node- and edge-removal attacks were investigated in the past few years. In particular, it was found that redundant edges, which are not included in any of the maximum matchings, can be rewired or re-directed so as to possibly enlarge a maximum matching such that the needed number of driver nodes is reduced \cite{Xu2014CCDC,Hou2013ISDEA}.

Although the correlation between network topology and network controllability has been investigated, there is no prominent theoretical indicator or performance index that can well describe the controllability robustness with this measure. Under different attack methods, the controllability robustness behaves differently. The essence of different attack methods leads to different definitions of \textit{importance} of nodes (or edges). Generally, degree and betweenness are commonly used measures for importance \cite{Pu2012PA}. In \cite{Liu2012PO}, a control centrality is defined to measure the importance of nodes, discovering that the upstream (or downstream) neighbors of a node are usually more (or less) important than the node itself. Interestingly, it was recently found that the existence of special motifs such as rings and chains is beneficial for enhancing the controllability robustness \cite{Lou2018TCASI,Chen2019TCASII,Lou2019R}.

On the other hand, in machine learning, deep neural networks have shown powerful capability in performing classification and regression tasks. Compared to the canonical neural networks that involve human-defined features, deep neural networks entirely ignore these requirements, thus lower down the risk of being misled by possibly biased human interference. For example, given an image classification task, a deep neural network works based only on the results of processing the raw image pixels, rather than human-defined features such as colors and textures. Convolutional neural network (CNN) is a kind of effective deep neural network. In a CNN, the convolutional layer consists of a set of learnable filters (called kernels) \cite{Schmidhuber2015NN}, and through a forward pass the well-trained filters are convolved with the input image maps. This structure is able to automatically analyze inner features of a dataset without human interference. Successful real-world applications of CNNs include text recognition and classification \cite{Wang2012ICPR,Lai2015AAAI, Zhang2015NIPS}, speech recognition \cite{Abdel2012ICASSP,Abdel2014TASLP}, and question answering \cite{Yin2016TACL,Qiu2015IJCAI} in natural language processing; image and video classifications \cite{Krizhevsky2012NIPS,Karpathy2014CVPR}, face recognition and detection \cite{Li2015CVPR}, and person re-identification \cite{Ahmed2015CVPR,Zhou2017CVPR} for performing vision tasks. Applications of CNNs have also been extended to biomedical image segmentation \cite{Ronneberger2015MICCAI}, patient-specific electrocardiogram classification \cite{Kiranyaz2015TBE}, etc.

\begin{figure*}[htbp]
	\begin{subfigure}{.25\textwidth}
		\centering
		\includegraphics[width=.7\linewidth]{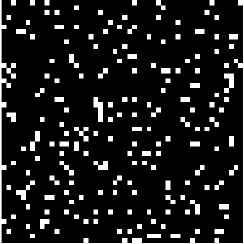}
		\caption{unweighted ER}
		\label{fig:fig1a}
	\end{subfigure}%
	\begin{subfigure}{.25\textwidth}
		\centering
		\includegraphics[width=.7\linewidth]{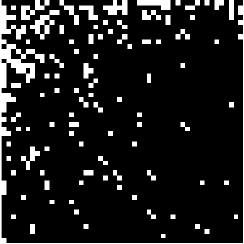}
		\caption{unweighted SF}
		\label{fig:fig1b}
	\end{subfigure}%
	\begin{subfigure}{.25\textwidth}
		\centering
		\includegraphics[width=.7\linewidth]{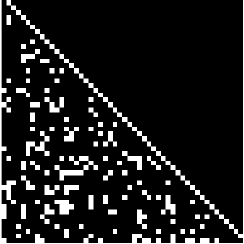}
		\caption{unweighted QSN}
		\label{fig:fig1c}
	\end{subfigure}%
	\begin{subfigure}{.25\textwidth}
		\centering
		\includegraphics[width=.7\linewidth]{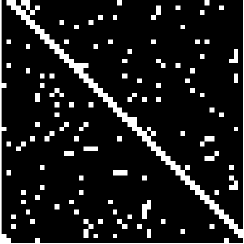}
		\caption{unweighted SW}
		\label{fig:fig1d}
	\end{subfigure}
	\begin{subfigure}{.25\textwidth}
		\centering
		\includegraphics[width=.7\linewidth]{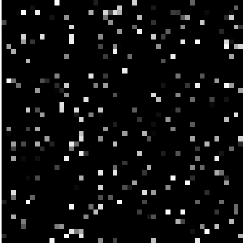}
		\caption{weighted ER}
		\label{fig:fig1e}
	\end{subfigure}%
	\begin{subfigure}{.25\textwidth}
		\centering
		\includegraphics[width=.7\linewidth]{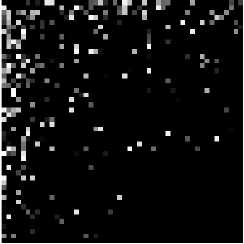}
		\caption{weighted SF}
		\label{fig:fig1f}
	\end{subfigure}%
	\begin{subfigure}{.25\textwidth}
		\centering
		\includegraphics[width=.7\linewidth]{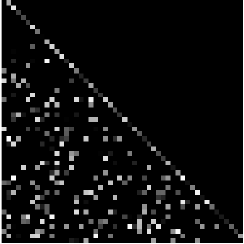}
		\caption{weighted QSN}
		\label{fig:fig1g}
	\end{subfigure}%
	\begin{subfigure}{.25\textwidth}
		\centering
		\includegraphics[width=.7\linewidth]{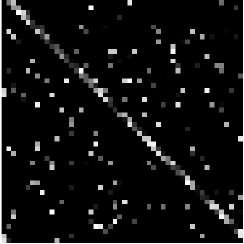}
		\caption{weighted SW}
		\label{fig:fig1h}
	\end{subfigure}
	\caption{An example of adjacency matrix converting images for both weighted and unweighted images. The network size $N=50$ with average degree $\langle k\rangle=5$. In each image, a black pixel represents a 0 element in the adjacency matrix; a white pixel represents a 1 element in the unweighted network. For an weighted network, the non-zero elements are normalized $\in(0,1]$ to form a gray-scale image. }
	\label{fig:fig_im}
\end{figure*}

In the literature, the measure of network controllability is quantified by the number of external controllers needed to recover or to retain the controllability after the occurrence of a malicious attack. The measure of the network controllability robustness is quantified by a sequence of values that record the remaining controllability of the network after a sequence of attacks. Usually, the controllability robustness is determined by attack simulation, which however is computationally time consuming. In this paper, a method to predict the controllability robustness using a CNN is proposed, based on the observations that there is no clear correlation between the topological features and the controllability robustness of a network, and the CNN technique has been successful in image processing without human intervention while a network adjacency matrix can be represented as a gray-scale image. Specifically, for a given adjacency matrix with 0-1 elements, representing an unweighted network, it is first converted to a black-white image; while for a weighted network with real-valued elements in the adjacency matrix, a gray-scale image is plotted. Examples of the converted images for both weighted and unweighted images are shown in Fig. \ref{fig:fig_im}, where different network topologies show distinguished patterns. The resulting images are then processed (for either training or testing) by a CNN, where the input is the raw image (adjacency matrix) and the output is referred to as a \textit{controllability curve} against the probability of attacks. Given an $N$-node network, a controllability curve is used to indicate the controllability of the remaining network after $i$ ($i=1,2,\ldots,N-1$) nodes (or edges) were removed. No feature of the adjacency matrix is assumed to have any correlation with the controllability robustness. Given a sufficiently large number of training samples, the CNN can be well trained for the purpose of prediction. Each sample (either for training or testing) consists of an input-output pair, i.e., a pair of an adjacency matrix and a controllability curve. Compared to the traditional network attack simulations, the proposed method costs significantly less time to obtain an accurate controllability robustness measure of a given network, of any size and any type, with low overheads in training. The performance of the proposed framework is also very encouraging, as further discussed in the experiments section below.

The rest part of the paper is organized as follows. Section \ref{sec:nc} reviews the network controllability and controllability robustness against various attacks. Section \ref{sec:cnn} introduces the convolutional neural network used in this work. In Section \ref{sec:exp}, experimental study is performed and analyzed. Finally, Section \ref{sec:con} concludes the investigation.

\section{Network Controllability}
\label{sec:nc}

Given a linear time-invariant (LTI) networked system described by $\dot{{\bf x}}=A{\bf x}+B{\bf u}$, where $A$ and $B$ are constant matrices of compatible dimensions, the system is \textit{state controllable} if and only if the controllability matrix $[B\ AB\ A^2B\ \cdots A^{N-1}B]$ has a full row-rank, where $N$ is the dimension of $A$. The concept of \textit{structural controllability} is a slight generalization dealing with two parameterized matrices $A$ and $B$, in which the parameters characterize the structure of the underlying networked system. If there are specific parameter values that can ensure the LTI system described by the two parameterized matrices be state controllable, then the underlying networked system is structurally controllable.

The network controllability is measured by the density of the driver nodes, $n_D$, defined by
\begin{equation}\label{eq:nd}
	n_D\equiv N_D/N\,,
\end{equation}
where $N_D$ is the number of driver nodes needed to retain the network controllability after the occurrence of an attack to the network, and $N$ is the current network size, which does not change during an edge-removal attack but would be reduced by a node-removal attack. Under this measure, the smaller the $n_D$ value is, the more robust the network controllability will be.

Recall that a network is sparse if the number of edges $M$ (or the number of nonzero elements of the adjacency matrix) is much less than the possible maximum number of edges, $M_{max}$. For a directed network, $M_{max}$ can be calculated by
\begin{equation}\label{eq:mmax}
	M_{max}=N\cdot (N-1)\,.
\end{equation}
Practically, if $M/M_{max}\leq 0.05$, then it is considered as a sparse network.

For state controllability, if the adjacency matrix $A$ of the network is sparse, the number $N_D$ of driver nodes can be calculated by \cite{Yuan2013NC}
\begin{equation}\label{eq:ec}
	N_D=\text{max}\{1, N-\text{rank}(A)\}.
\end{equation}
If it has a full rank, then the number of driver nodes is $N_D=1$; otherwise, $N_D=N-\text{rank}(A)$ diver nodes are needed, which should be properly assigned (to fill the deficiency of the matrix rank).

As for structural controllability, according to the \textit{minimum inputs theorem} \cite{Liu2011N}, when a maximum matching is found, the number of driver nodes $N_D$ is determined by the number of unmatched nodes, i.e.,
\begin{equation}\label{eq:sc}
	N_D=\text{max}\{1, N-|E^*|\},
\end{equation}
where $|E^*|$ is the cardinal number of elements in the maximum matching $E^*$. In a directed network, a matching is a set of edges that do not share common start or end nodes. A maximum matching is a matching that contains the largest possible number of edges, which cannot be further extended. A node is matched if it is the end of an edge in the matching; otherwise, it is unmatched. A perfect matching is a matching that matches all nodes in the network. If a network has a perfect matching, then the number of driver nodes is $N_D=1$ and the driver node can be any node; otherwise, the needed $N_D=N-|E^*|$ control inputs should be put at those unmatched nodes.

The robustness of both state and structural controllabilities is reflected by the value of $n_D$ calculated according to Eq. (\ref{eq:nd}) and recorded after each node or edge is removed. In this paper, for brevity, only node-removal attacks are discussed.

\section{Convolutional Neural Networks}
\label{sec:cnn}

The CNN to be used for controllability performance prediction consists of an embedding layer, several groups of convolutional layers, and some fully connected layers. Rectified linear unit (ReLU) is employed as the activation function and max pooling is used to reduce the dimensions of datasets.

\begin{figure*}[htbp]	
	\begin{center}
		\includegraphics[width=.85\textwidth]{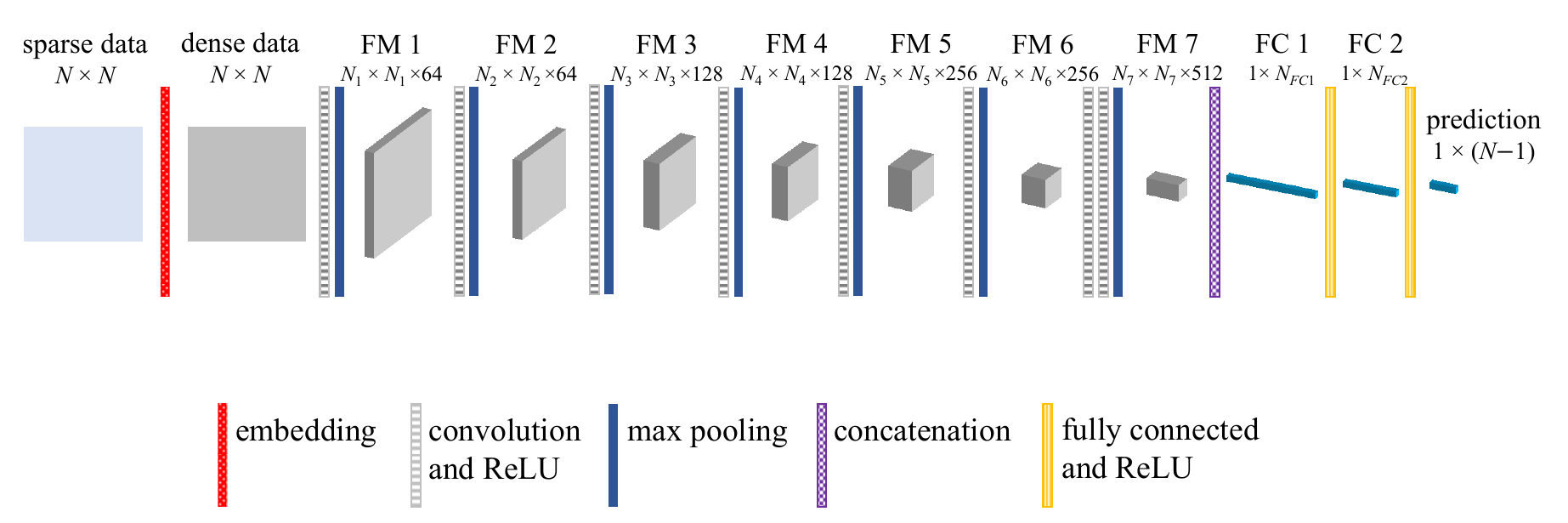}
		\caption{The architecture of the CNN used for controllability robustness prediction, where FM is an abbreviation for \textit{feature map}, and FC for \textit{fully connected}. The data size $N_i=\ceil{N/(i+1)}$, for $i=1,2,\ldots,7$. The concatenation layer reshapes the matrix to a vector, from FM $7$ to FC $1$, i.e., $N_{FC1}=N_7\times N_7\times 512$. $N_{FC2}$ is a hyperparameter and $N_{FC2}\in (N_{FC1},N-1)$. Set $N_{FC2}=4096$ for $N=800$, $1000$, and $1200$, in this paper.}
		\label{fig:cnn}
	\end{center}
\end{figure*}

\begin{table}[htbp]
	\centering
	\caption{Parameter settings of the seven groups of convolutional layers.}
	\begin{tabular}{|c|r|c|c|c|} \hline
		Group & \multicolumn{1}{c|}{Layer} & \begin{tabular}[c]{@{}c@{}}Kernel\\ size\end{tabular}
              & Stride & \begin{tabular}[c]{@{}c@{}}Output\\ channel\end{tabular} \\ \hline
		\multirow{2}{*}{Group 1} & Conv7-64 & 7x7 & 1 & 64 \\ \cline{2-5} & Max2 & 2x2 & 2 & 64 \\ \hline
		\multirow{2}{*}{Group 2} & Conv5-64 & 5x5 & 1 & 64 \\ \cline{2-5} & Max2 & 2x2 & 2 & 64 \\ \hline
		\multirow{2}{*}{Group 3} & Conv3-128 & 3x3 & 1 & 128 \\ \cline{2-5} & Max2 & 2x2 & 2 & 128 \\ \hline
		\multirow{2}{*}{Group 4} & Conv3-128 & 3x3 & 1 & 128 \\ \cline{2-5} & Max2 & 2x2 & 2 & 128 \\ \hline
		\multirow{2}{*}{Group 5} & Conv3-256 & 3x3 & 1 & 256 \\ \cline{2-5} & Max2 & 2x2 & 2 & 256 \\ \hline
		\multirow{2}{*}{Group 6} & Conv3-256 & 3x3 & 1 & 256 \\ \cline{2-5} & Max2 & 2x2 & 2 & 256 \\ \hline
		\multirow{2}{*}{Group 7} & Conv3-512 & 3x3 & 1 & 512 \\ \cline{2-5} & Max2 & 2x2 & 2 & 512 \\ \hline
	\end{tabular} \label{tab:cnn_para}
\end{table}

The framework of CNN is shown in Fig. \ref{fig:cnn}. The detailed parameter settings of the $7$ groups of convolutional layers are given in Table \ref{tab:cnn_para}. Here, the CNN architecture follows the Visual Geometry Group (VGG)\footnote{\url{http://www.robots.ox.ac.uk/~vgg/}}) architecture \cite{Simonyan2014arXiv}. Compared with the original CNN architectures, the VGG architecture incorporates a smaller kernel size and a greater network depth. Thus, in this architecture, most kernels have small sizes ($2\times2$ or $3\times3$). The convolution stride is set to $1$ and the pooling stride is set to $2$.

Given the input as an $N\times N$ sparse matrix with less than $5\%$ of non-zero elements but more than $95\%$ zeros. If the network is unweighted then further assume that the input to the CNN is a sparse one-hot matrix, namely a 0-1 matrix that contains less than $M_{max}\times 5\%$ 1-elements. According to \cite{Mikolov2013arXiv}, however, deep neural network performs poorly on this type of data, and thus the input data will be converted to normal dense data via the embedding layer for better performances. An embedding layer is a randomly generated normalized matrix, which has a size $N\times N$ here. The converted matrix $A'=AD$ becomes a dense one, where $A$ is the original sparse (one-hot) matrix and $D$ is the embedding matrix.

As shown in Fig. \ref{fig:cnn}, following the preprocessing embedding layer, there are $7$ feature map (FM) processing layers, denoted as FM $1$ to FM $7$. Each FM processing layer consists of a convolution layer, a ReLU, and a max pooling layer.

Convolutional layers are used in the hidden layers because they can more efficiently deal with large-scale datasets (e.g., $1000\times1000$ pixel images), while fully-connected layers are more suitable for small-sized datasets such as MNIST\footnote{\url{http://yann.lecun.com/exdb/mnist/}} images (an MNIST image has $28\times28$ pixels). The size of a real complex network generally varies from hundreds (e.g., there are $279$ non-pharyngeal neurons in the sensory system of the worm \textit{C. elegans} \cite{Yan2017NAT}) to billions (e.g., $5.5$ billions of pages in the World Wide Web\footnote{\url{https://www.worldwidewebsize.com/} (Accessed: 20 May, 2020)}). In Section \ref{sec:exp}, the network size for simulation will be set to $1000$, and thus the convolutional layers are employed instead of fully-connected layers. Compared to the fully-connected layers, using the convolutional layers can significantly reduce the number of parameters. Meanwhile, the number of FM groups will be set to $7$, since the input size is around $1000\times1000$ in the following experiments. The number of FM groups should be set to be greater for a larger input dataset, and vice versa.

ReLU provides a commonly-used activation function that defines the non-negative part of a set of inputs. It has been proved performing the best on $2D$ data \cite{Glorot2011ICAIS}. The formation of ReLU to be used here is $f(x)=\text{max}\{0,x\}$. The output of each hidden layer, i.e., a multiplication of weights, is summed up and rectified by a ReLU for the next layer.

Pooling layers reduce the dimensions of the datasets for the input of the following layer. There are two commonly-used pooling methods, namely the max pooling and average pooling. The former uses the maximum value (a greater value represents a brighter pixel in a gray-scale image) within the current window, while the latter uses the average value. Max pooling is useful when the background of the image is dark. Since the interest here is only in the lighter pixels, the max pooling is adopted.

Following the $7$ convolutional groups, there are two fully-connected layers. The last feature map, FM $7$, is concatenated to FC $1$, and thus the input size of FC $1$ is equal to the number of elements of the FM $7$ output, which is $N_7\times N_7\times512$. The size of FC 2 is a hyper-parameter that should be set to be within the range $N_{FC2}\in (N_{FC1},N-1)$.

The loss function used is equal to the mean-squared error between the predicted vector (referred to as a controllability curve) and the true vector, as follows:
\begin{equation}\label{eq:lf}
	\mathcal{L} = \frac{1}{N-1} \sum_{i=1}^{N-1}|| pv_i-tv_i ||,
\end{equation}
where $||\cdot||$ is the Euclidean norm, $pv_i$ is the $i$th element of the predicted curve, and $tv_i$ is the $i$th element of the true controllability curve. The training process for the CNN aims to minimize the loss function (\ref{eq:lf}). Source codes of this work are available for the public\footnote{\url{https://fylou.github.io/sourcecode.html}}.

\section{Experimental Study}
\label{sec:exp}

Four representative directed networks, each has weighted or unweighted edges, are examined. They are Erd{\"{o}}s-R{\'e}nyi random graph (ER) \cite{Erdos1964RG}, generic scale-free (SF) network \cite{Pu2012PA,Goh2001PRL,Sorrentino2007CH}, \textit{q}-snapback network (QSN) \cite{Lou2018TCASI}, and Newman--Watts small-world (SW) network \cite{Newman1999PLA}.

Specifically, the edges in the ER network are added completely at random, where the direction of each edge is evenly-randomly assigned. The edges of SF network are generated according to their weights defined by
\begin{equation}\label{eq:sf}
	w_{i}=(i+\theta)^{-\sigma}, i=1,2,\ldots,N,
\end{equation}
where $\sigma\in[0,1)$ and $\theta\ll N$. Two nodes $i$ and $j$ ($i\neq j$, $i,j=1,2,...,N$) are randomly picked with a probability proportional to the weights $w_i$ and $w_j$, respectively. Then, an edge $A_{ij}$ from $i$ to $j$ is added (if the two nodes are already connected, do nothing). The resulting SF network has a power-law distribution $k^{-\gamma}$, where $k$ is the degree variable, with constant $\gamma=1+\frac{1}{\sigma}$, which is independent of $\theta$. In this paper, $\sigma$ is set to 0.999, such that the power-law distribution has a reasonable scaling exponent $\gamma=2.001$. The QSN has one layer, $r_{q}$, generated with a backbone chain and multiple snapback edges created for each node $i=r_{q}+1,\, r_{q}+2,\, \ldots, N$, which connects backward to the previously-appeared nodes $i-l\times r_{q}$ ($l=1,2,\ldots, \floor{i/r_{q}}$), with the same probability $q\in[0,1]$ \cite{Lou2018TCASI}. The SW network is initiated by a directed ring with $K=2$ nearest-neighbors connected, i.e., a node $i$ is connected to nodes $i-1$, $i+1$, $i-2$ and $i+2$, via edges $A_{i-1,i}$, $A_{i,i+1}$, $A_{i-2,i}$ and $A_{i,i+2}$. Then, adding or removing edges, until the predefined number of edges is reached.

For each network topology, unweighted networks are studied in Subsection \ref{subsec:uw} and weighted networks are studied in Subsection \ref{subsec:wt}. The adjacency matrix is represented by
\begin{equation}\label{eq:uw}
A_{i,j}=
	\begin{cases}
		0, & \textrm{no edge connecting } i \textrm{ to } j,\\
		e, & \textrm{otherwise}.\\
	\end{cases}
\end{equation}
where $e=\textrm{rand}(0,1]$ representing a random real value for a weighted network; $e=1$ for a unweighted network. Examples of images that are converted from weighted or unweighted networks are shown in Fig. \ref{fig:fig_im}.

Resizing (upsampling or subsampling) is commonly used for CNNs to process inputs with different sizes. However, therein the application scenario is different. For example, portrait photos can be properly resized, without disturbing the task of object recognition. However, a pixel here represents an edge in the network, and thus resizing will change the original topology, thereby misleading the task of prediction. There are a few works that deal with different network sizes by using the same CNN, with additional assumptions. For example, in \cite{Niepert2016ICML}, to generate an input sequence for a network of any size, a sequence of nodes that covers large parts of the network should be searched first, and moreover the local neighborhood structure has to be assumed. In this paper, CNNs are used to process both raw and complete structures of complex networks, without any assumption or prior knowledge on network structural features, and thus the input size is fixed to be same as the size of the input network. Due to space limitation, only node-removal attacks are considered in this paper.

The node-removal attack methods include: 1) random attack (RA) that removes randomly-selected nodes; 2) targeted betweenness-based attack (TBA) that removes nodes with maximum betweenness; and 3) targeted degree-based attack (TDA) that removes nodes with maximum node degree. For TBA and TDA, when two or more nodes have the same maximum value of betweenness or degree, one node is randomly picked to remove.

The experiments are run on a PC with 64-bit operation system, installed Intel i7-6700 (3.4 GHz) CPU, GeForce GTX 1080 Ti GPU. The CNN is implemented in TensorFlow\footnote{\url{https://www.tensorflow.org/}} that is a Python-friendly open source library.

\subsection{Controllability Curve Prediction}
\label{subsec:uw}

First, unweighted networks of size $N=1000$ are considered. For each network topology, four average (out-)degree values are set, namely $\langle k\rangle=2$, $\langle k\rangle=5$, $\langle k\rangle=8$, and $\langle k\rangle=10$. Since the networks are unweighted, only the structural controllability (see Eq. (\ref{eq:sc})) is discussed here. In Subsection \ref{subsec:wt}, state controllability (see Eq. (\ref{eq:ec})) is measured for the weighted networks discussed therein. The reason is that the structural controllability is independent of the edge weights, while the state controllability depend on them.

For CNN training and testing, with each of the three attack methods, there are $1000$ randomly generated instances for each network configuration, amongst which $800$ instances are used for training, $100$ instances for cross validation, and the other $100$ instances for testing. Taking RA for example, there are $4(\textrm{topologies})\times4(\langle k\rangle\textrm{ values})\times800$ instances used for CNN training; $4\times4\times100$ instances used for cross validation; and $4\times4\times100$ instances used for testing. The number of training epochs is empirically set to $10$ in these simulations. For each training epoch, the CNN is trained by the full set of training data in a shuffled order. To prevent over-fitting, after running an epoch, the trained CNN is evaluated by the full set of validation data. The best-performing structure on the validation data is then selected as the optimum structure. The Adam optimizer \cite{Kingma2014arXiv} is used to optimize the CNN, i.e., to minimize the loss function defined in Eq. (\ref{eq:lf}). The learning rate of Adam is set to $0.005$. The first momentum, the second momentum and the $\epsilon$ of Adam follow the recommended settings, which are $0.9$, $0.999$ and $1\times10^{-8}$, respectively.

\begin{figure*}[htbp]
	\begin{subfigure}{.25\textwidth}
		\centering
		\includegraphics[width=\linewidth]{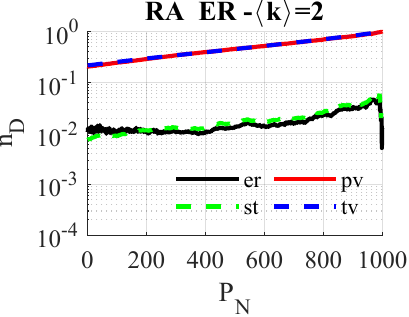}
		\caption{}
		\label{fig:fig3a}
	\end{subfigure}%
	\begin{subfigure}{.25\textwidth}
		\centering
		\includegraphics[width=\linewidth]{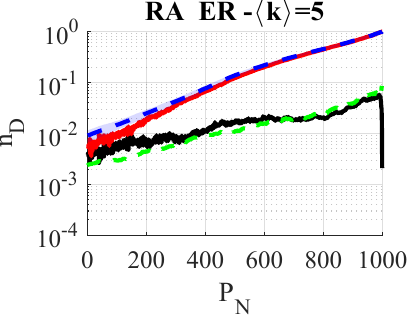}
		\caption{}
		\label{fig:fig3b}
	\end{subfigure}%
	\begin{subfigure}{.25\textwidth}
		\centering
		\includegraphics[width=\linewidth]{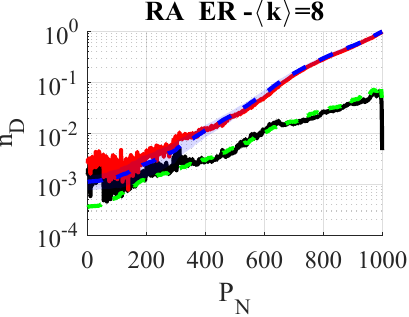}
		\caption{}
		\label{fig:fig3c}
	\end{subfigure}%
	\begin{subfigure}{.25\textwidth}
		\centering
		\includegraphics[width=\linewidth]{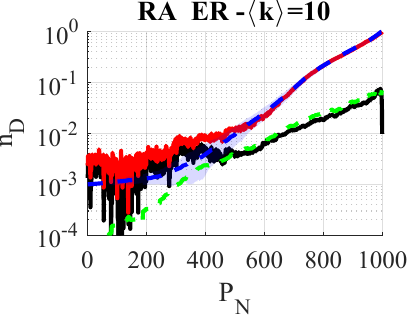}
		\caption{}
		\label{fig:fig3d}
	\end{subfigure}
	\begin{subfigure}{.25\textwidth}
		\centering
		\includegraphics[width=\linewidth]{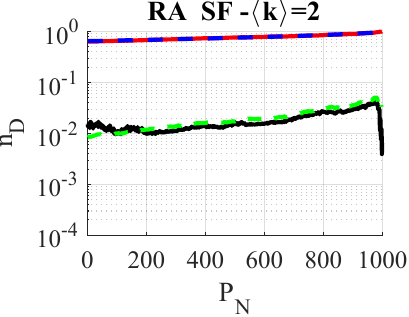}
		\caption{}
		\label{fig:fig3e}
	\end{subfigure}%
	\begin{subfigure}{.25\textwidth}
		\centering
		\includegraphics[width=\linewidth]{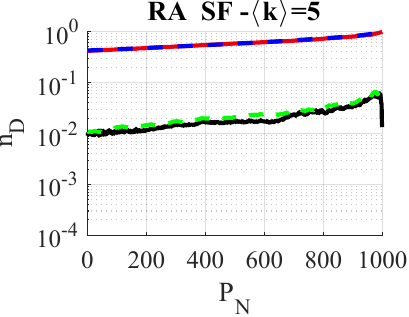}
		\caption{}
		\label{fig:fig3f}
	\end{subfigure}%
	\begin{subfigure}{.25\textwidth}
		\centering
		\includegraphics[width=\linewidth]{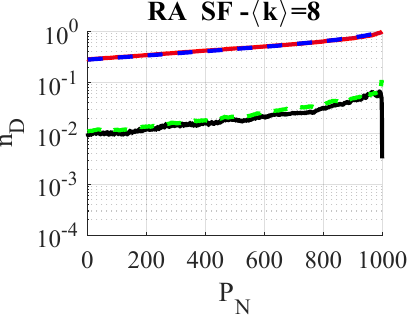}
		\caption{}
		\label{fig:fig3g}
	\end{subfigure}%
	\begin{subfigure}{.25\textwidth}
		\centering
		\includegraphics[width=\linewidth]{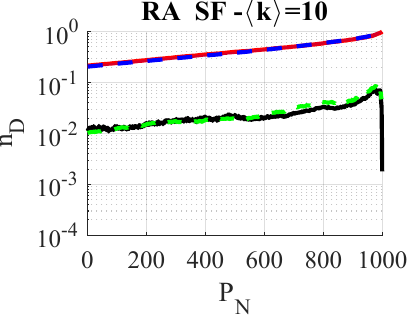}
		\caption{}
		\label{fig:fig3h}
	\end{subfigure}
	\begin{subfigure}{.25\textwidth}
		\centering
		\includegraphics[width=\linewidth]{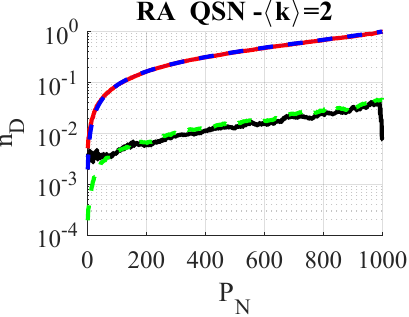}
		\caption{}
		\label{fig:fig3i}
	\end{subfigure}%
	\begin{subfigure}{.25\textwidth}
		\centering
		\includegraphics[width=\linewidth]{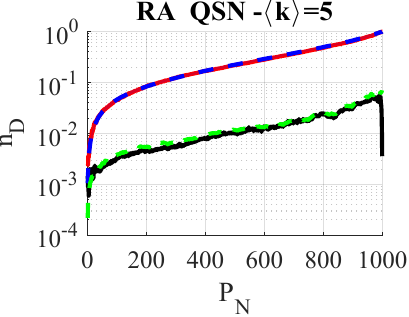}
		\caption{}
		\label{fig:fig3j}
	\end{subfigure}%
	\begin{subfigure}{.25\textwidth}
		\centering
		\includegraphics[width=\linewidth]{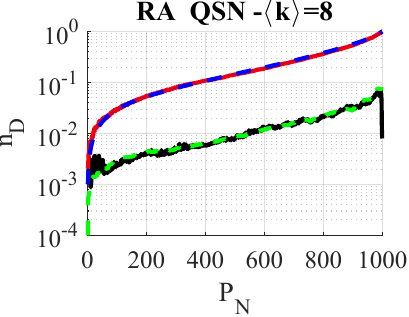}
		\caption{}
		\label{fig:fig3k}
	\end{subfigure}%
	\begin{subfigure}{.25\textwidth}
		\centering
		\includegraphics[width=\linewidth]{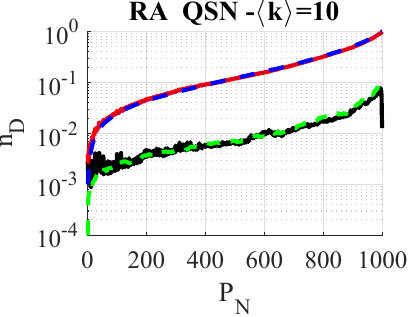}
		\caption{}
		\label{fig:fig3l}
	\end{subfigure}
	\begin{subfigure}{.25\textwidth}
		\centering
		\includegraphics[width=\linewidth]{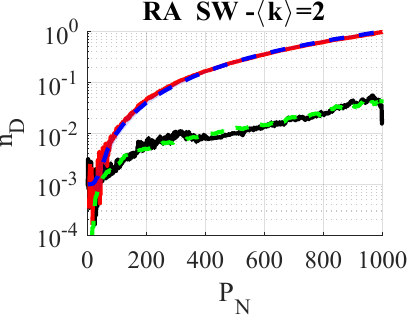}
		\caption{}
		\label{fig:fig3m}
	\end{subfigure}%
	\begin{subfigure}{.25\textwidth}
		\centering
		\includegraphics[width=\linewidth]{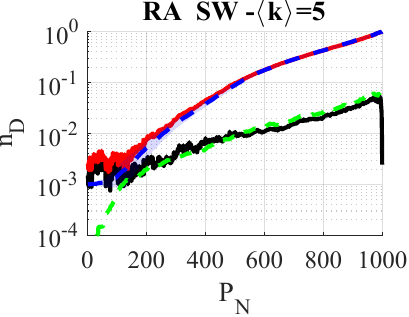}
		\caption{}
		\label{fig:fig3n}
	\end{subfigure}%
	\begin{subfigure}{.25\textwidth}
		\centering
		\includegraphics[width=\linewidth]{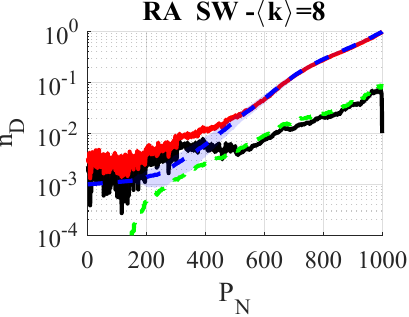}
		\caption{}
		\label{fig:fig3o}
	\end{subfigure}%
	\begin{subfigure}{.25\textwidth}
		\centering
		\includegraphics[width=\linewidth]{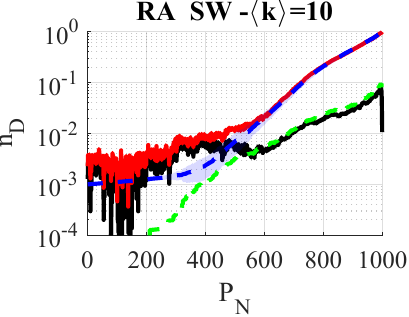}
		\caption{}
		\label{fig:fig3p}
	\end{subfigure}
	\caption{[Color online] Results of CNN controllability curve prediction under random attacks. $P_N$ represents the number of nodes having been removed from the network; $n_D$ is calculated by Eqs. (\ref{eq:nd}) and (\ref{eq:sc}).}
	\label{fig:fig_ra}
\end{figure*}

Figure \ref{fig:fig_ra} shows the testing results on RA. The results of TBA and TDA are shown in Figs. S1 and S2 of the supplementary information (SI)\footnote{\url{https://fylou.github.io/pdf/pcrsi.pdf}}. In these figures, results of the same topology with different average degrees are aligned in a row. The average degree increases from left to right on each row. The red curve shows the predicted value ($pv$) given by the CNN; the dotted blue line represents the true value ($tv$) obtained from simulations; the black curve shows the error ($er$) between the prediction and the true value; and the green dotted line represents the standard deviation of the true value ($st$), used as reference. For each curve of $pv$, $tv$, and $er$, the shaded shadow in the same color shows the standard deviation, while the standard deviation curve $st$ does not contain such a shadow.

It can be seen from these figures that different network topologies show different controllability curves against the same attacks. For QSN and SW, since there is a directed-chain in QSN and a connected global loop in SW, the initial number of needed external controllers is 1, hence the controllability curves start from $10^{-3}$. In contrast, there is no such a structure in ER or SF, thus their curves start from much higher than $10^{-3}$. From left to right, as the average degree increases, 1) the positions of all curves become lower, which means less controllers are needed, and the controllability consequently becomes better, throughout the entire attack process; 2) the initial positions of ER and SF curves become lower, but for QSN and SW, their initial positions remain the same at $10^{-3}$.

Generally, both $er$ and $st$ increase as the number of removed nodes increases. The two curves $er$ and $st$ generally match each other, which means that the prediction error and the standard deviation of the testing data roughly have the same magnitude. Even in some cases, $er$ is less than $st$ mostly throughout the attack process (e.g., shown by Fig. \ref{fig:fig_ra}(f)), when the $tv$ and $pv$ values approach 1, both $er$ and $st$ curves drop drastically.

The predicted controllability curves become rugged for the two homogeneous random graphs, ER and SW, in the initial stage of RA. As $\langle k\rangle$ increases, the ruggedness of controllability curves increase as well. This is because the topological characteristics (the unknown features abstracted by CNN itself) of the homogeneous networks can only be mildly influenced by the change of average degree. In the latter stage of RA, the predicted controllability curves match the true curves smoothly. In contrast, SF is a heterogeneous network while QSN has a uniform degree distribution; both topologies have stronger topological characteristics and thus the predicted curves and true curves are matched smoothly. As shown in the converted gray-scale image examples (see Fig. \ref{fig:fig_im}), SF images have a strong characteristic, where the light pixels (representing edges) are converged in the upper-left corner. Except for the black upper-right triangle and the backbone-chain, the rest light pixels in QSN are uniformly randomly distributed. Similarly, except for the backbone directed triangles, the rest light pixels in SW are also uniformly randomly distributed. There are substantial numbers of light pixels in ER, QSN, and SW, which are uniformly randomly distributed in the images.

Figures S1 and S2 in SI show the results on TBA and TDA, respectively. Similarly, the predicted curves and the true curves are smoothly matched for both SF and QSN, while the predicted curves are rugged for both ER and SW in the initial stage of an attack, especially as $\langle k\rangle$ increases. It is worth mentioning that in Figs. S1(c), S1(d), S1(n), S1(o), S1(p), and Figs. S2(c), S2(d), S2(m), S2(n), S2(o), S2(p), the $pv$ curves cannot well match the $tv$ curves at the very beginning of the process, i.e., at the initial positions. This means that the controllability robustness of homogeneous networks are more difficult to predict than that on heterogeneous networks and QSN, which have stronger topological features. Meanwhile, the controllability performance prediction on TBA and TDA is also slightly more difficult than that on RA. Although the $er$ values in the rugged part is relatively small (compared to the other parts of the curve), they are nevertheless greater than the $st$ values.

Note that, during CNN training, networks with different topologies and different average degrees are trained together. The controllability performance prediction is not specified for any particular network topology with a specified average degree. As for an attack method, it can also be integrated into the training data. However, here the focus is on predicting the controllability curves, rather than predicting the attack method, hence different attack methods are trained and tested separately.

As shown in Figs. \ref{fig:fig_ra}, and Figs. S1 and S2 in SI that the value of $er$ is visibly as small as the standard deviation of the testing data, which include $100$ testing samples. In Table \ref{tab:uw_std}, the mean error is compared to the mean standard deviation of the testing data under RA. The mean error $\bar{er}$ is further averaged over the $999$ data points, i.e., $\bar{er}$ equals the average value of a whole ${er}$ curve. The mean standard deviation $\bar{\sigma}$ is the mean of the standard deviations calculated on $P_N=i$ ($i=1,2,\ldots,N-1$, $N=1000$), i.e., the average value of a whole ${st}$ curve. As can be seen from the table, the $\bar{er}$ values are less than or equal to the $\bar{\sigma}$ values in most cases, indicating that the overall prediction error is very small, thus the prediction of the CNN is very precise. The comparisons under TBA and TDA are presented in Table S1 of SI.

\begin{table}[htbp]
	\centering
	\caption{The mean error of prediction vs. the mean standard deviation of the testing data for the unweighted networks under RA.}
	\begin{tabular}{|c|c|c|c|c|c|c|}
		\hline
		\multicolumn{3}{|l|}{}& $\langle k\rangle=2$ & $\langle k\rangle=5$ & $\langle k\rangle=8$ & $\langle k\rangle=10$ \\ \hline
		\multirow{8}{*}{RA} & \multirow{2}{*}{ER} & $\bar{er}$  & 0.017 & 0.017 & 0.013 & 0.011 \\
     \cline{3-7} && $\bar{\sigma}$ & 0.019 & 0.017 & 0.015 & 0.013 \\ \cline{2-7}
		& \multirow{2}{*}{SF}  & $\bar{er}$ & 0.018 & 0.020 & 0.023 & 0.024 \\ \cline{3-7}
		&& $\bar{\sigma}$ & 0.021 & 0.024 & 0.026 & 0.026 \\ \cline{2-7}
		& \multirow{2}{*}{QSN} & $\bar{er}$ & 0.015 & 0.014 & 0.013 & 0.012 \\ \cline{3-7}
		&& $\bar{\sigma}$ & 0.018 & 0.016 & 0.014 & 0.013 \\ \cline{2-7}
		& \multirow{2}{*}{SW}  & $\bar{er}$ & 0.015 & 0.013 & 0.013 & 0.011 \\ \cline{3-7}
		&& $\bar{\sigma}$ & 0.015 & 0.015 & 0.014 & 0.012 \\ \hline
	\end{tabular}
	\label{tab:uw_std}
\end{table}

\subsection{Extension of Experiments}
\label{subsec:wt}

The previous subsection has demonstrated that the (structural) controllability robustness performance of various networks, with different average degrees under different attacks, can be well predicted. Here, the experiments are further extended to predict state controllability curves of weighted networks, with network size set to $N=800$, $1000$, and $1200$, respectively. For each network size setting, a new CNN is trained. The average degree is fixed to $\langle k\rangle=10$. The CNN structure and configurations remain the same as in the experiments reported in Subsection \ref{subsec:uw}, where only the input and output sizes are modified. 

\begin{figure*}[htbp]
	\begin{subfigure}{.25\textwidth}
		\centering
		\includegraphics[width=\linewidth]{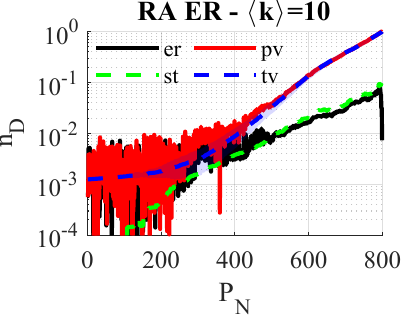}
		\caption{weighted ER $N=800$}
		\label{fig:fig4a}
	\end{subfigure}%
	\begin{subfigure}{.25\textwidth}
		\centering
		\includegraphics[width=\linewidth]{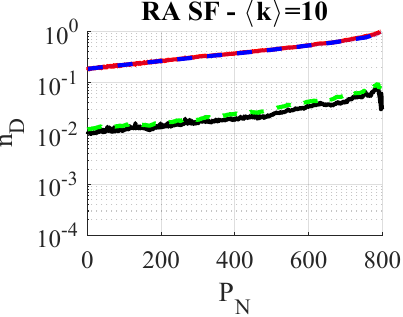}
		\caption{weighted SF $N=800$}
		\label{fig:fig4b}
	\end{subfigure}%
	\begin{subfigure}{.25\textwidth}
		\centering
		\includegraphics[width=\linewidth]{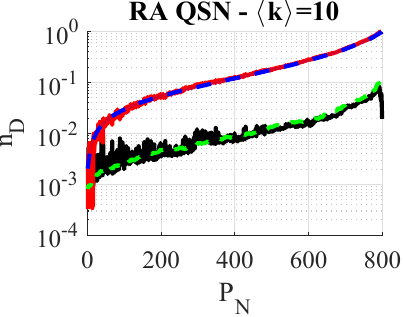}
		\caption{weighted QSN $N=800$}
		\label{fig:fig4c}
	\end{subfigure}%
	\begin{subfigure}{.25\textwidth}
		\centering
		\includegraphics[width=\linewidth]{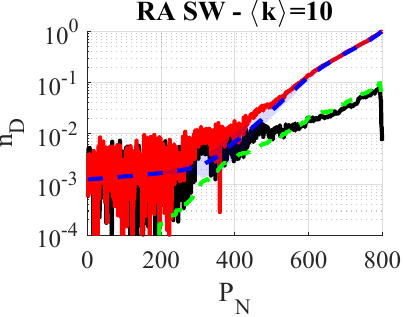}
		\caption{weighted SW $N=800$}
		\label{fig:fig4d}
	\end{subfigure}
	\begin{subfigure}{.25\textwidth}
		\centering
		\includegraphics[width=\linewidth]{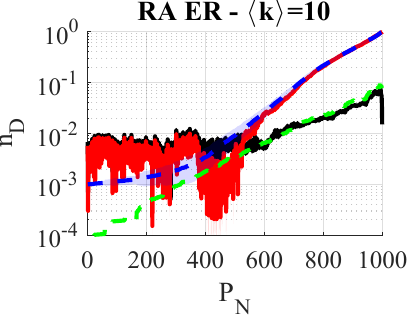}
		\caption{weighted ER $N=1000$}
		\label{fig:fig4e}
	\end{subfigure}%
	\begin{subfigure}{.25\textwidth}
		\centering
		\includegraphics[width=\linewidth]{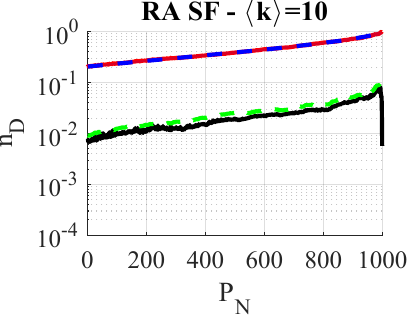}
		\caption{weighted SF $N=1000$}
		\label{fig:fig4f}
	\end{subfigure}%
	\begin{subfigure}{.25\textwidth}
		\centering
		\includegraphics[width=\linewidth]{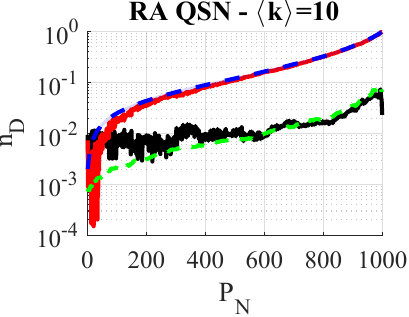}
		\caption{weighted QSN $N=1000$}
		\label{fig:fig4g}
	\end{subfigure}%
	\begin{subfigure}{.25\textwidth}
		\centering
		\includegraphics[width=\linewidth]{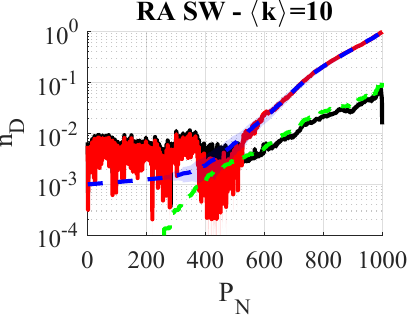}
		\caption{weighted SW $N=1000$}
		\label{fig:fig4h}
	\end{subfigure}
	\begin{subfigure}{.25\textwidth}
		\centering
		\includegraphics[width=\linewidth]{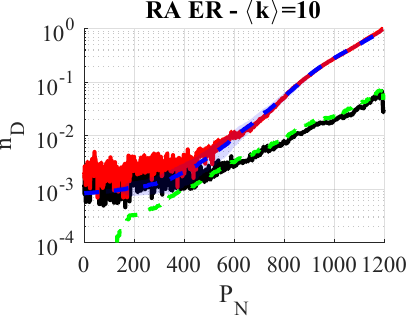}
		\caption{weighted ER $N=1200$}
		\label{fig:fig4i}
	\end{subfigure}%
	\begin{subfigure}{.25\textwidth}
		\centering
		\includegraphics[width=\linewidth]{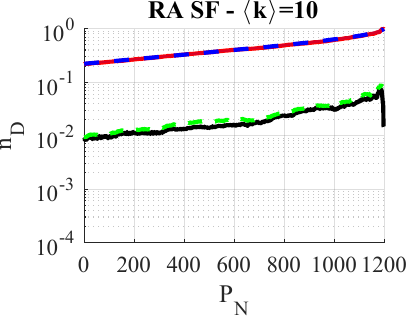}
		\caption{weighted SF $N=1200$}
		\label{fig:fig4j}
	\end{subfigure}%
	\begin{subfigure}{.25\textwidth}
		\centering
		\includegraphics[width=\linewidth]{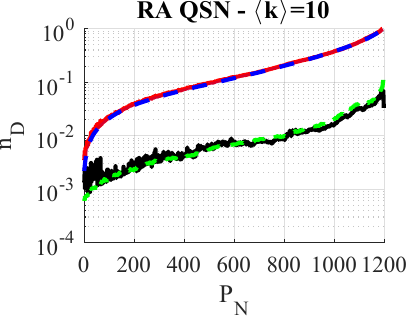}
		\caption{weighted QSN $N=1200$}
		\label{fig:fig4k}
	\end{subfigure}%
	\begin{subfigure}{.25\textwidth}
		\centering
		\includegraphics[width=\linewidth]{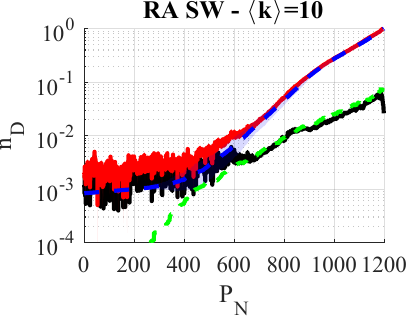}
		\caption{weighted SW $N=1200$}
		\label{fig:fig4l}
	\end{subfigure}
	\caption{[Color online] Results of CNN controllability curve prediction on weighted networks under different attacks. $P_N$ represents the number of nodes having been removed from the network; $n_D$ is calculated by Eqs. (\ref{eq:nd}) and (\ref{eq:ec}).}
	\label{fig:fig_weighted}
\end{figure*}

The experimental results are shown in Fig. \ref{fig:fig_weighted} and Table \ref{tab:wt_std}. Fig. \ref{fig:fig_weighted} shows that the prediction error $er$ remains low when the network size is varied from $800$, to $1000$, then to $1200$. Once again, in Table \ref{tab:wt_std} the comparison between the mean error $\bar{er}$ of the results and the mean standard deviation $\bar{\sigma}$ of the testing data shows very high precision of the prediction.

\begin{table}[htbp]
	\centering
	\caption{The mean error of prediction vs. the mean standard deviation of the testing data
		for the weighted networks.}
	\begin{tabular}{|c|c|c|c|c|c|}
		\hline
		\multicolumn{3}{|l|}{} & $N=800$ & $N=1000$ & $N=1200$ \\ \hline
		\multirow{8}{*}{RA} & \multirow{2}{*}{ER}  & $\bar{er}$  & 0.013 & 0.014  & 0.010 \\ \cline{3-6}
		&& $\bar{\sigma}$ & 0.014 & 0.013  & 0.011  \\ \cline{2-6}
		& \multirow{2}{*}{SF}  & $\bar{er}$  & 0.025 & 0.022  & 0.021 \\ \cline{3-6}
		&& $\bar{\sigma}$ & 0.031 & 0.028  & 0.025  \\ \cline{2-6}
		& \multirow{2}{*}{QSN} & $\bar{er}$  & 0.014 & 0.014  & 0.011 \\ \cline{3-6}
		&& $\bar{\sigma}$ & 0.016 & 0.014  & 0.012  \\ \cline{2-6}
		& \multirow{2}{*}{SW}  & $\bar{er}$  & 0.014 & 0.012  & 0.010 \\ \cline{3-6}
		&& $\bar{\sigma}$ & 0.013 & 0.012  & 0.011  \\ \hline
	\end{tabular}
	\label{tab:wt_std}
\end{table}

\begin{figure*}[htbp]
	\begin{subfigure}{.25\textwidth}
		\centering
		\includegraphics[width=\linewidth]{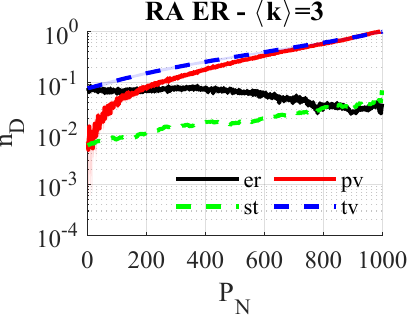}
		\caption{}
		\label{fig:fig5a}
	\end{subfigure}%
	\begin{subfigure}{.25\textwidth}
		\centering
		\includegraphics[width=\linewidth]{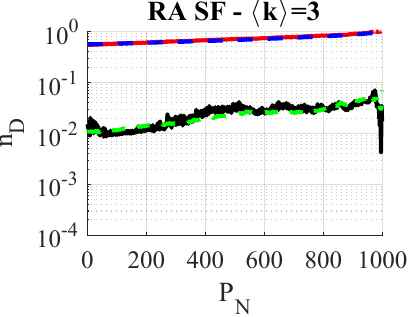}
		\caption{}
		\label{fig:fig5b}
	\end{subfigure}%
	\begin{subfigure}{.25\textwidth}
		\centering
		\includegraphics[width=\linewidth]{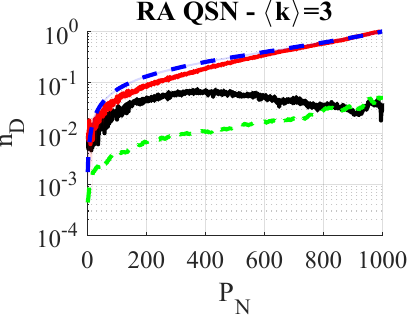}
		\caption{}
		\label{fig:fig5c}
	\end{subfigure}%
	\begin{subfigure}{.25\textwidth}
		\centering
		\includegraphics[width=\linewidth]{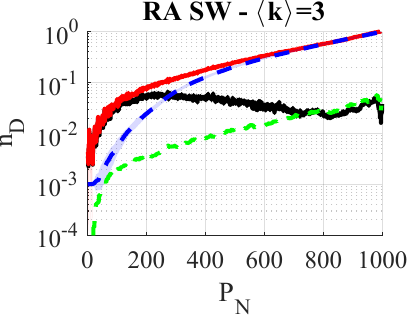}
		\caption{}
		\label{fig:fig5d}
	\end{subfigure}
	\begin{subfigure}{.25\textwidth}
		\centering
		\includegraphics[width=\linewidth]{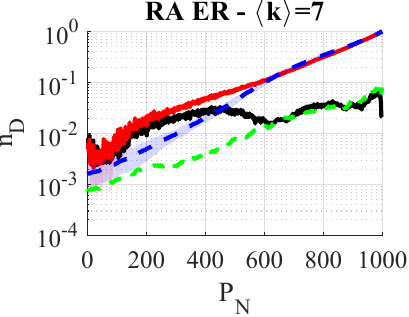}
		\caption{}
		\label{fig:fig5e}
	\end{subfigure}%
	\begin{subfigure}{.25\textwidth}
		\centering
		\includegraphics[width=\linewidth]{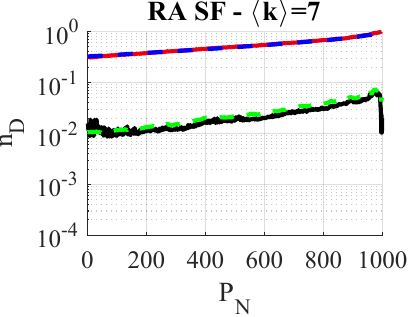}
		\caption{}
		\label{fig:fig5f}
	\end{subfigure}%
	\begin{subfigure}{.25\textwidth}
		\centering
		\includegraphics[width=\linewidth]{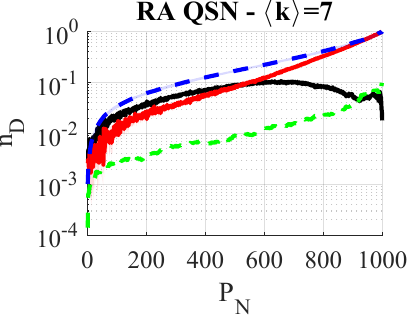}
		\caption{}
		\label{fig:fig5g}
	\end{subfigure}%
	\begin{subfigure}{.25\textwidth}
		\centering
		\includegraphics[width=\linewidth]{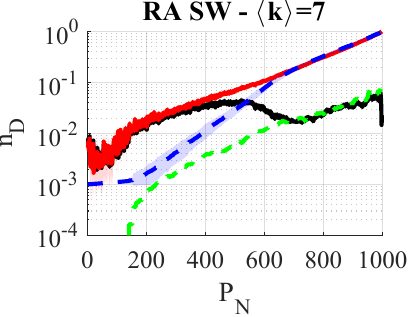}
		\caption{}
		\label{fig:fig5h}
	\end{subfigure}
	\caption{[Color online] Results of CNN controllability curve prediction on unweighted networks with $\langle k\rangle=3$ and $7$ respectively, under random attacks. The CNN is trained in the same way as that in Fig. \ref{fig:fig_ra} ($\langle k\rangle=2$, $5$, $8$, and $10$). $P_N$ represents the number of nodes removed from the network; $n_D$ is calculated by Eqs. (\ref{eq:nd}) and (\ref{eq:sc}).}
	\label{fig:fig_k37}
\end{figure*}

The extensibility of the trained CNN is investigated, when the networks of the training and testing sets have different average degrees. The CNN trained in Subsection \ref{subsec:uw} (for RA) is reused here without further training, where unweighted networks with average degrees $\langle k\rangle=2$, $5$, $8$, and $10$ respectively, are employed. Then, the CNN is used to predict the controllability robustness for the networks with $\langle k\rangle=3$ and $7$ respectively. The network size is $N=1000$ and the structural controllability is calculated. Compared to Fig. \ref{fig:fig_ra}, where the training and testing data are drawn from the same distribution, when the training and testing data have different distributions, the prediction performances of ER (Figs. \ref{fig:fig_k37}(a) and (e)), QSN (Figs. \ref{fig:fig_k37}(c) and (g)), and SW (Figs. \ref{fig:fig_k37}(d) and (h)) are clearly worse. In contrast, the controllability robustness of SF (Figs. \ref{fig:fig_k37}(b) and (f)) can be well predicted. Table \ref{tab:k37} shows that, for ER, QSN, and SW, the mean errors of prediction $\bar{er}$ are about 2 to 4 times greater than the mean standard deviations $\bar{\sigma}$ of the testing data; while for SF, the prediction reaches a very low mean error that is close to $\bar{\sigma}$.

It is remarked that, when the training and testing data are drawn from different distributions, a more suitable machine learning technique might be transfer learning \cite{Pan2009TKDE}.

\begin{table}[htbp]
	\centering
	\caption{The mean error of prediction vs. the mean standard deviation, where the training and testing data have different average degrees.}
	\begin{tabular}{|c|c|c|c|c|}
		\hline
		\multicolumn{3}{|c|}{} & $\langle k\rangle=3$ & $\langle k\rangle=7$ \\ \hline
		\multirow{8}{*}{RA} & \multirow{2}{*}{ER}  & er  & 0.058 & 0.023 \\ \cline{3-5}
		&                      & $\bar{\sigma}$ & 0.020 & 0.016 \\ \cline{2-5}
		& \multirow{2}{*}{SF}  & $\bar{er}$ & 0.025 & 0.022 \\ \cline{3-5}
		&                      & $\bar{\sigma}$ & 0.024 & 0.027 \\ \cline{2-5}
		& \multirow{2}{*}{QSN} & $\bar{er}$ & 0.044 & 0.063 \\ \cline{3-5}
		&                      & $\bar{\sigma}$ & 0.016 & 0.015 \\ \cline{2-5}
		& \multirow{2}{*}{SW}  & $\bar{er}$ & 0.037 & 0.026 \\ \cline{3-5}
		&                      & $\bar{\sigma}$ & 0.015 & 0.014 \\ \hline
	\end{tabular}
	\label{tab:k37}
\end{table}

\begin{figure*}[htbp]
	\begin{subfigure}{.25\textwidth}
		\centering
		\includegraphics[width=\linewidth]{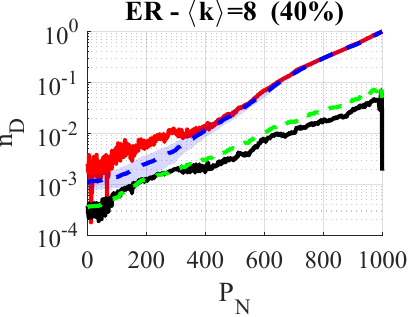}
		\caption{}
	\end{subfigure}%
	\begin{subfigure}{.25\textwidth}
		\centering
		\includegraphics[width=\linewidth]{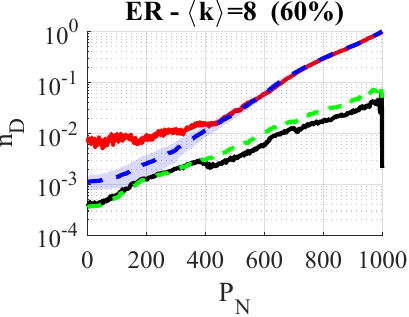}
		\caption{}
	\end{subfigure}%
	\begin{subfigure}{.25\textwidth}
		\centering
		\includegraphics[width=\linewidth]{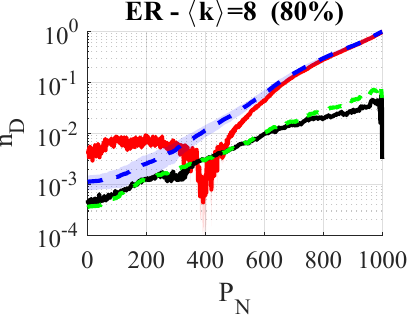}
		\caption{}
	\end{subfigure}%
	\begin{subfigure}{.25\textwidth}
		\centering
		\includegraphics[width=\linewidth]{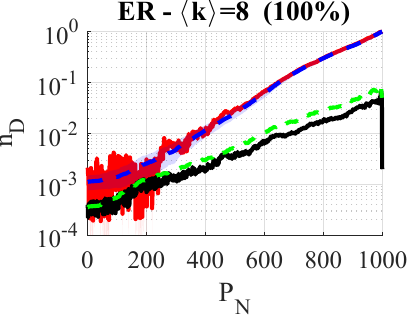}
		\caption{}
	\end{subfigure}	
	\begin{subfigure}{.25\textwidth}
		\centering
		\includegraphics[width=\linewidth]{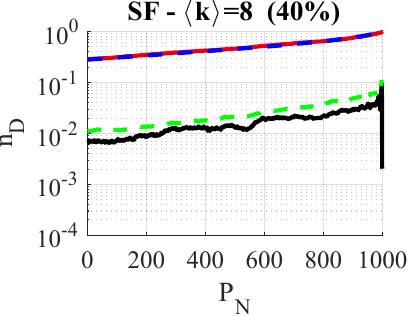}
		\caption{}
	\end{subfigure}%
	\begin{subfigure}{.25\textwidth}
		\centering
		\includegraphics[width=\linewidth]{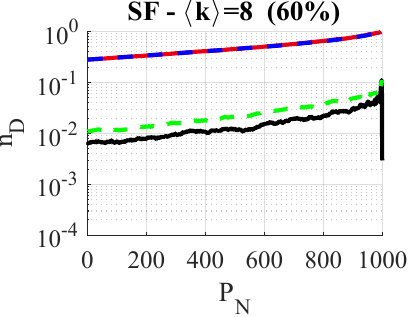}
		\caption{}
	\end{subfigure}%
	\begin{subfigure}{.25\textwidth}
		\centering
		\includegraphics[width=\linewidth]{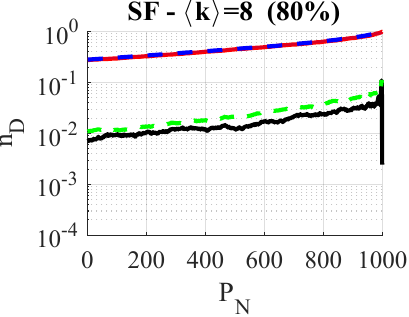}
		\caption{}
	\end{subfigure}%
	\begin{subfigure}{.25\textwidth}
		\centering
		\includegraphics[width=\linewidth]{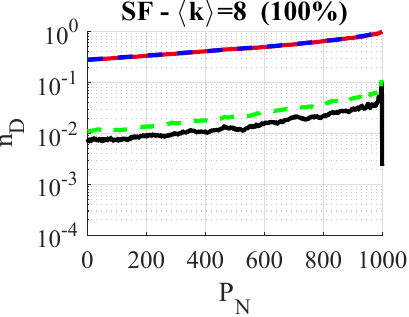}
		\caption{}
	\end{subfigure}
	\begin{subfigure}{.25\textwidth}
		\centering
		\includegraphics[width=\linewidth]{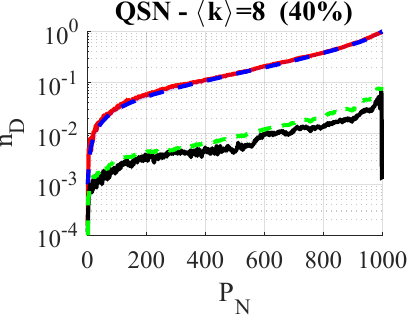}
		\caption{}
	\end{subfigure}%
	\begin{subfigure}{.25\textwidth}
		\centering
		\includegraphics[width=\linewidth]{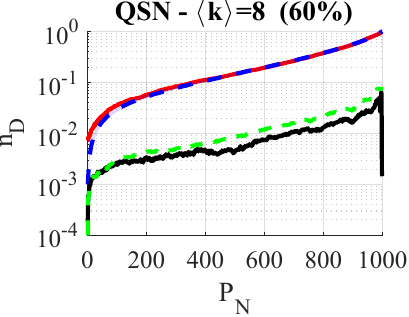}
		\caption{}
	\end{subfigure}%
	\begin{subfigure}{.25\textwidth}
		\centering
		\includegraphics[width=\linewidth]{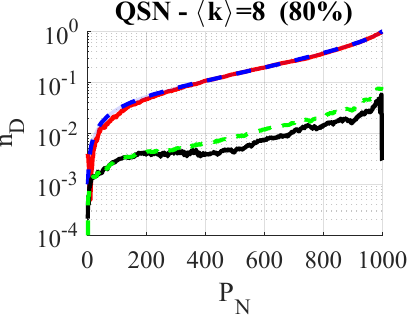}
		\caption{}
	\end{subfigure}%
	\begin{subfigure}{.25\textwidth}
		\centering
		\includegraphics[width=\linewidth]{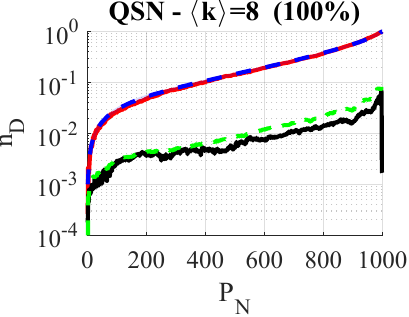}
		\caption{}
	\end{subfigure}	
	\begin{subfigure}{.25\textwidth}
		\centering
		\includegraphics[width=\linewidth]{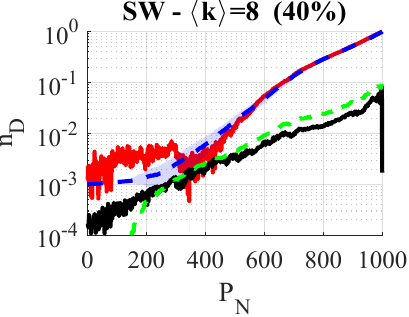}
		\caption{}
	\end{subfigure}%
	\begin{subfigure}{.25\textwidth}
		\centering
		\includegraphics[width=\linewidth]{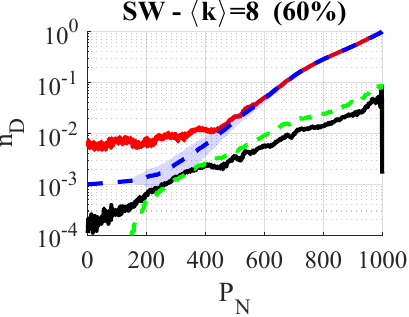}
		\caption{}
	\end{subfigure}%
	\begin{subfigure}{.25\textwidth}
		\centering
		\includegraphics[width=\linewidth]{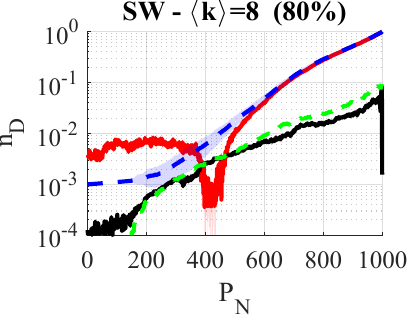}
		\caption{}
	\end{subfigure}%
	\begin{subfigure}{.25\textwidth}
		\centering
		\includegraphics[width=\linewidth]{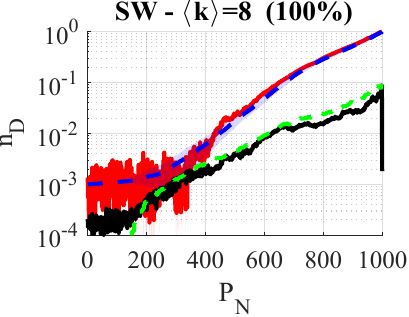}
		\caption{}
	\end{subfigure}
	\caption{[Color online] Results of CNN controllability curve prediction under random attacks. The size of training data is set to $40\%$, $60\%$, $80\%$, and $100\%$ of the data size used in the paper, where $100\%$ represents a training set of $800$ instances. The average degree is set to $\langle k\rangle=8$.}
	\label{fig:size}
\end{figure*}

Finally, consider different sizes of training data in the settings of Subsection \ref{subsec:uw}, where $800$ training instances are employed for each topology with a given $\langle k\rangle$. Next, the size of training data is reduced to $40\%$, $60\%$, and $80\%$ of that, respectively. Meanwhile, the data sizes of cross validation set and testing set remain the same.

Figure \ref{fig:size} shows the prediction results when $\langle k\rangle=8$. It can be seen that a smaller size of training data does not influence the prediction performance for SF and QSN, but clearly influences the results for ER and SW. This is because SF and QSN have strong topological characteristics. A full set of prediction results for the four network topologies with $\langle k\rangle=2$, $5$, $8$, and $10$, are shown in Figs. S3--S6, respectively. The average error comparison is presented in Table S2. It is revealed that, if a network has strong topological characteristics (e.g., SF and QSN), or if its average degree is low (e.g., $\langle k\rangle=2$), then a small training data size can be used; otherwise, a setting as in Subsection \ref{subsec:uw} is recommended.

\subsection{Computational Costs}
\label{subsec:cost}

\begin{figure}[htbp]	
	\begin{center}
		\includegraphics[width=.48\textwidth]{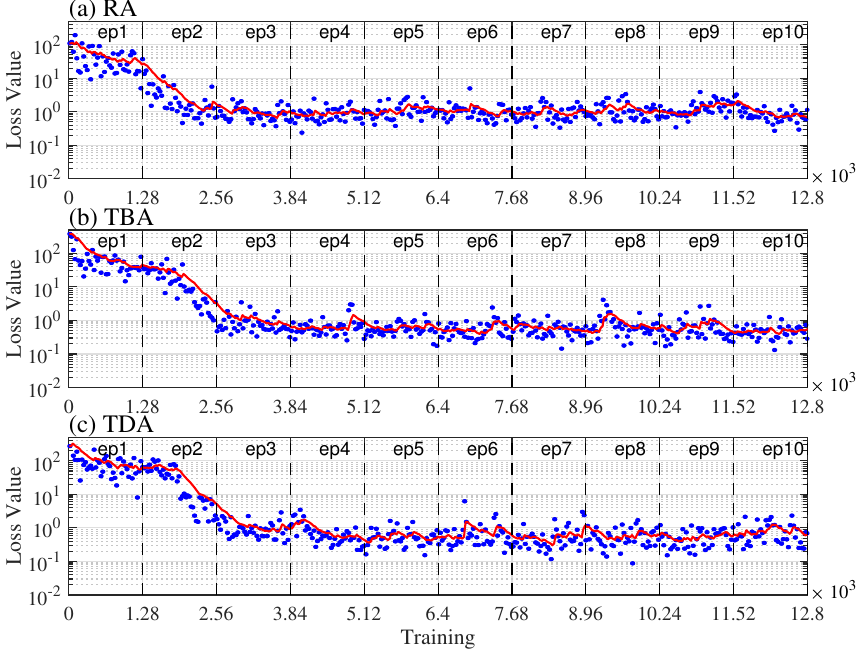}
		\caption{The convergence process of loss value calculated by Eq. (\ref{eq:lf}) during the CNN training of (a) RA; (b) TBA; and (c) TDA. The horizontal axis `Training' represents the number of instances trained. The red curve represents the exponentially weighted averages (with $\beta=0.9$) for reference. There are $10$ epochs in total. In each epoch, $4\times4\times800=12800$ instances are trained.}
		\label{fig:loss}
	\end{center}
\end{figure}

Given an $N$-node network without self-loops and multiple links, there could be $N\cdot(N-1)$ directed edges at most. For sparse networks considered here, with a small average degree $\langle k\rangle=a$, there are $aN$ edges in total, with ${N\cdot(N-1)\choose aN}$ possible combinations. This shows the impossibility of exhaustively enumerating all possible networks with fixed size and average degree. Further, with a given adjacency matrix at hand, the simulation to measure its controllability robustness is non-trivial. The simulation process should be iterated for $N-1$ times. At each time step $t$, when $t$ nodes have been removed, the major computational cost includes: 1) to find the node with the maximum betweenness or degree among the present $N-t$ nodes; 2) to calculate the current controllability, using Eq. (\ref{eq:ec}) or Eq. (\ref{eq:sc}), both are time-consuming in traditional simulation.

Empirically, given a PC with configuration as introduced above, the elapsed time is about 90 seconds for a QSN with $N=1000$ and $\langle k\rangle=10$ under the attack by TDA, to collect a controllability robustness curve by simulation. In contrast, it needs less than 0.2 seconds to get a controllability robustness predicted curve by a well-trained CNN. The training overhead is also low, which costs less than 5 hours to train a CNN using $4\times4\times800$ samples. The convergence process of loss values is plotted in Fig. \ref{fig:loss}. An example of run time comparison is available with source codes\footnote{\url{https://fylou.github.io/sourcecode.html}}.

In summary, it is not only effective, but also efficient to train a CNN and then use it to predict the network controllability robustness.

\section{Conclusions}
\label{sec:con}

Network controllability robustness, which reflects how well a networked system can maintain a full control under various malicious attacks, can be measured via attack simulations, which returns the true values of the controllability robustness but is computationally costly and time consuming in conventional approaches. In this paper, the convolutional neural network is employed to predict the controllability robustness, which returns encouraging results, in the sense that the prediction error is of the same magnitude as the standard deviation of the data samples, with attractively low computational costs. Convolutional neural network is adopted in this study, for the following reasons: 1) no network topological features have been found and reported in the literature, to provide a clear correlation with the controllability robustness measure; 2) the adjacency matrix of a complex network can be processed as a gray-scale image; 3) convolutional neural network technique has been successfully applied to image processing where no clear feature-label correlations are given in advance. The present study has confirmed that convolutional neural network is indeed an excellent choice for predicting the complex network controllability robustness. In this study, extensive experimental simulations were carried out to predict the controllability robustness of four typical network topologies, in both weighted and unweighted settings, with different network sizes and average degrees. For all cases, the convolutional neural network tool is proved to provide accurate and reliable predictions of the controllability robustness for all simulated networks under different kinds of malicious attacks. Future research aims to develop analytic methods for in-depth studies of the controllability robustness and other related issues in general complex networks.


\ifCLASSOPTIONcaptionsoff
 \newpage
\fi

\bibliographystyle{IEEEtran}

\begin{thebibliography}{10}
	\providecommand{\url}[1]{#1}
	\csname url@samestyle\endcsname
	\providecommand{\newblock}{\relax}
	\providecommand{\bibinfo}[2]{#2}
	\providecommand{\BIBentrySTDinterwordspacing}{\spaceskip=0pt\relax}
	\providecommand{\BIBentryALTinterwordstretchfactor}{4}
	\providecommand{\BIBentryALTinterwordspacing}{\spaceskip=\fontdimen2\font plus
		\BIBentryALTinterwordstretchfactor\fontdimen3\font minus
		\fontdimen4\font\relax}
	\providecommand{\BIBforeignlanguage}[2]{{%
			\expandafter\ifx\csname l@#1\endcsname\relax
			\typeout{** WARNING: IEEEtran.bst: No hyphenation pattern has been}%
			\typeout{** loaded for the language `#1'. Using the pattern for}%
			\typeout{** the default language instead.}%
			\else
			\language=\csname l@#1\endcsname
			\fi
			#2}}
	\providecommand{\BIBdecl}{\relax}
	\BIBdecl
	
	\bibitem{Barabasi2016NS}
	A.-L. Barab{\'a}si, \emph{Network Science}.\hskip 1em plus 0.5em minus
	0.4em\relax Cambridge University Press, 2016.
	
	\bibitem{Newman2010N}
	M.~E. Newman, \emph{Networks: An Introduction}.\hskip 1em plus 0.5em minus
	0.4em\relax Oxford University Press, 2010.
	
	\bibitem{Chen2014Book}
	G.~Chen, X.~Wang, and X.~Li, \emph{Fundamentals of Complex Networks: Models,
		Structures and Dynamics}, 2nd~ed.\hskip 1em plus 0.5em minus 0.4em\relax John
	Wiley \& Sons, 2014.
	
	\bibitem{Liu2011N}
	Y.-Y. Liu, J.-J. Slotine, and A.-L. Barab{\'a}si, ``Controllability of complex
	networks,'' \emph{Nature}, vol. 473, no. 7346, pp. 167--173, 2011.
	
	\bibitem{Yuan2013NC}
	Z.~Z. Yuan, C.~Zhao, Z.~R. Di, W.-X. Wang, and Y.-C. Lai, ``Exact
	controllability of complex networks,'' \emph{Nature Communications}, vol.~4,
	p. 2447, 2013.
	
	\bibitem{Posfai2013SR}
	M.~P{\'o}sfai, Y.-Y. Liu, J.-J. Slotine, and A.-L. Barab{\'a}si, ``Effect of
	correlations on network controllability,'' \emph{Scientific Reports}, vol.~3,
	p. 1067, 2013.
	
	\bibitem{Menichetti2014PRL}
	G.~Menichetti, L.~Dall'Asta, and G.~Bianconi, ``Network controllability is
	determined by the density of low in-degree and out-degree nodes,''
	\emph{Physical Review Letters}, vol. 113, no.~7, p. 078701, 2014.
	
	\bibitem{Motter15CHAOS}
	A.~E. Motter, ``Networkcontrology,'' \emph{Chaos: An Interdisciplinary Journal
		of Nonlinear Science}, vol.~25, no.~9, p. 097621, 2015.
	
	\bibitem{Wang2016AUTO}
	L.~Wang, X.~Wang, G.~Chen, and W.~K.~S. Tang, ``Controllability of networked
	mimo systems,'' \emph{Automatica}, vol.~69, pp. 405--409, 2016.
	
	\bibitem{Liu2016RMP}
	Y.-Y. Liu and A.-L. Barab{\'a}si, ``Control principles of complex systems,''
	\emph{Review of Modern Physics}, vol.~88, no.~3, p. 035006, 2016.
	
	\bibitem{Wang2017RSPTA}
	L.~Wang, X.~Wang, and G.~Chen, ``Controllability of networked
	higher-dimensional systems with one-dimensional communication channels,''
	\emph{Royal Society Philosophical Transactions A}, vol. 375, no. 2088, p.
	20160215, 2017.
	
	\bibitem{Wang20L17SR}
	L.-Z. Wang, Y.-Z. Chen, W.-X. Wang, and Y.-C. Lai, ``Physical controllability
	of complex networks,'' \emph{Scientific Reports}, vol.~7, p. 40198, 2017.
	
	\bibitem{Zhang2017TAC}
	Y.~Zhang and T.~Zhou, ``Controllability analysis for a networked dynamic system
	with autonomous subsystems,'' \emph{IEEE Transactions on Automatic Control},
	vol.~62, no.~7, pp. 3408--3415, 2016.
	
	\bibitem{Wu2018JNS}
	E.~Wu-Yan, R.~F. Betzel, E.~Tang, S.~Gu, F.~Pasqualetti, and D.~S. Bassett,
	``Benchmarking measures of network controllability on canonical graph
	models,'' \emph{Journal of Nonlinear Science}, pp. 1--39, 2018.
	
	\bibitem{Hao2018IJRNC}
	Y.~Hao, Z.~Duan, and G.~Chen, ``Further on the controllability of networked
	{MIMO LTI} systems,'' \emph{International Journal of Robust and Nonlinear
		Control}, vol.~28, no.~5, pp. 1778--1788, 2018.
	
	\bibitem{Xiang2019CSM}
	L.~Xiang, F.~Chen, W.~Ren, and G.~Chen, ``Advances in network
	controllability,'' \emph{IEEE Circuits and Systems Magazine}, vol.~19, no.~2,
	pp. 8--32, 2019.
	
	\bibitem{Wen2020TSMC}
	G.~Wen, X.~Yu, W.~Yu, and J.~L{\"u}, ``Coordination and control of complex
	network systems with switching topologies: A survey,'' \emph{IEEE
		Transactions on Systems, Man, and Cybernetics: Systems}, 2020,
	doi:10.1109/TSMC.2019.2961753.
	
	\bibitem{Holme2002PRE}
	P.~Holme, B.~J. Kim, C.~N. Yoon, and S.~K. Han, ``Attack vulnerability of
	complex networks,'' \emph{Physical Review E}, vol.~65, no.~5, p. 056109,
	2002.
	
	\bibitem{Shargel2003PRL}
	B.~Shargel, H.~Sayama, I.~R. Epstein, and Y.~Bar-Yam, ``Optimization of
	robustness and connectivity in complex networks,'' \emph{Physical Review
		Letters}, vol.~90, no.~6, p. 068701, 2003.
	
	\bibitem{Schneider2011PNAS}
	C.~M. Schneider, A.~A. Moreira, J.~S. Andrade, S.~Havlin, and H.~J. Herrmann,
	``Mitigation of malicious attacks on networks,'' \emph{Proceedings of the
		National Academy of Sciences}, vol. 108, no.~10, pp. 3838--3841, 2011.
	
	\bibitem{Liu2012PO}
	Y.-Y. Liu, J.-J. Slotine, and A.-L. Barab{\'a}si, ``Control centrality and
	hierarchical structure in complex networks,'' \emph{PLOS ONE}, vol.~7, no.~9,
	p. e44459, 2012.
	
	\bibitem{BBBH13}
	A.~Bashan, Y.~Berezin, S.~Buldyrev, and S.~Havlin, ``The extreme vulnerability
	of interdependent spatially embedded networks,'' \emph{Nature Physics},
	vol.~9, pp. 667--672, 2013.
	
	\bibitem{Xiao2014CPB}
	Y.-D. Xiao, S.-Y. Lao, L.-L. Hou, and L.~Bai, ``Optimization of robustness of
	network controllability against malicious attacks,'' \emph{Chinese Physics
		B}, vol.~23, no.~11, p. 118902, 2014.
	
	\bibitem{Pu2012PA}
	C.-L. Pu, W.-J. Pei, and A.~Michaelson, ``Robustness analysis of network
	controllability,'' \emph{Physica A: Statistical Mechanics and its
		Applications}, vol. 391, no.~18, pp. 4420--4425, 2012.
	
	\bibitem{Nie2014PO}
	S.~Nie, X.~Wang, H.~Zhang, Q.~Li, and B.~Wang, ``Robustness of controllability
	for networks based on edge-attack,'' \emph{PLoS One}, vol.~9, no.~2, p.
	e89066, 2014.
	
	\bibitem{BPPSH10}
	S.~Buldyrev, R.~Parshani, G.~Paul, H.~Stanley, and S.~Havlin, ``Catastrophic
	cascade of failures in interdependent networks,'' \emph{Nature}, vol. 464, p.
	1025, 2010.
	
	\bibitem{Lou2018TCASI}
	Y.~Lou, L.~Wang, and G.~Chen, ``Toward stronger robustness of network
	controllability: {A} snapback network model,'' \emph{IEEE Transactions on
		Circuits and Systems I: Regular Papers}, vol.~65, no.~9, pp. 2983--2991,
	2018.
	
	\bibitem{Xu2014CCDC}
	J.~Xu, J.~Wang, H.~Zhao, and S.~Jia, ``Improving controllability of complex
	networks by rewiring links regularly,'' in \emph{Chinese Control and Decision
		Conference ({CCDC})}, 2014, pp. 642--645.
	
	\bibitem{Hou2013ISDEA}
	L.~Hou, S.~Lao, B.~Jiang, and L.~Bai, ``Enhancing complex network
	controllability by rewiring links,'' in \emph{International Conference on
		Intelligent System Design and Engineering Applications ({ISDEA})}.\hskip 1em
	plus 0.5em minus 0.4em\relax IEEE, 2013, pp. 709--711.
	
	\bibitem{Chen2019TCASII}
	G.~Chen, Y.~Lou, and L.~Wang, ``A comparative study on controllability
	robustness of complex networks,'' \emph{IEEE Transactions on Circuits and
		Systems II: Express Briefs}, vol.~66, no.~5, pp. 828--832, 2019.
	
	\bibitem{Lou2019R}
	Y.~Lou, L.~Wang, and G.~Chen, ``Enhancing controllability robustness of
	\textit{q}-snapback networks through redirecting edges,'' \emph{Research},
	vol. 2019, no. 7857534, 2019.
	
	\bibitem{Schmidhuber2015NN}
	J.~Schmidhuber, ``Deep learning in neural networks: An overview,'' \emph{Neural
		Networks}, vol.~61, pp. 85--117, 2015.
	
	\bibitem{Wang2012ICPR}
	T.~Wang, D.~J. Wu, A.~Coates, and A.~Y. Ng, ``End-to-end text recognition with
	convolutional neural networks,'' in \emph{International Conference on Pattern
		Recognition (ICPR 2012)}.\hskip 1em plus 0.5em minus 0.4em\relax IEEE, 2012,
	pp. 3304--3308.
	
	\bibitem{Lai2015AAAI}
	S.~Lai, L.~Xu, K.~Liu, and J.~Zhao, ``Recurrent convolutional neural networks
	for text classification,'' in \emph{AAAI Conference on Artificial
		Intelligence}, 2015, pp. 2267--2273.
	
	\bibitem{Zhang2015NIPS}
	X.~Zhang, J.~Zhao, and Y.~LeCun, ``Character-level convolutional networks for
	text classification,'' in \emph{Advances in Neural Information Processing
		Systems (NIPS 2015)}, 2015, pp. 649--657.
	
	\bibitem{Abdel2012ICASSP}
	O.~Abdel-Hamid, A.-r. Mohamed, H.~Jiang, and G.~Penn, ``Applying convolutional
	neural networks concepts to hybrid {NN-HMM} model for speech recognition,''
	in \emph{International Conference on Acoustics, Speech and Signal Processing
		(ICASSP)}.\hskip 1em plus 0.5em minus 0.4em\relax IEEE, 2012, pp. 4277--4280.
	
	\bibitem{Abdel2014TASLP}
	O.~Abdel-Hamid, A.-R. Mohamed, H.~Jiang, L.~Deng, G.~Penn, and D.~Yu,
	``Convolutional neural networks for speech recognition,'' \emph{IEEE/ACM
		Transactions on Audio, Speech, and Language Processing}, vol.~22, no.~10, pp.
	1533--1545, 2014.
	
	\bibitem{Yin2016TACL}
	W.~Yin, H.~Sch{\"u}tze, B.~Xiang, and B.~Zhou, ``ABCNN: Attention-based
	convolutional neural network for modeling sentence pairs,''
	\emph{Transactions of the Association for Computational Linguistics}, vol.~4,
	pp. 259--272, 2016.
	
	\bibitem{Qiu2015IJCAI}
	X.~Qiu and X.~Huang, ``Convolutional neural tensor network architecture for
	community-based question answering,'' in \emph{International Joint Conference
		on Artificial Intelligence}, 2015, pp. 1305--1311.
	
	\bibitem{Krizhevsky2012NIPS}
	A.~Krizhevsky, I.~Sutskever, and G.~E. Hinton, ``Imagenet classification with
	deep convolutional neural networks,'' in \emph{Advances in Neural Information
		Processing Systems (NIPS 2012)}, 2012, pp. 1097--1105.
	
	\bibitem{Karpathy2014CVPR}
	A.~Karpathy, G.~Toderici, S.~Shetty, T.~Leung, R.~Sukthankar, and L.~Fei-Fei,
	``Large-scale video classification with convolutional neural networks,'' in
	\emph{IEEE Conference on Computer Vision and Pattern Recognition (CVPR)},
	2014, pp. 1725--1732.
	
	\bibitem{Li2015CVPR}
	H.~Li, Z.~Lin, X.~Shen, J.~Brandt, and G.~Hua, ``A convolutional neural network
	cascade for face detection,'' in \emph{IEEE Conference on Computer Vision and
		Pattern Recognition (CVPR)}, 2015, pp. 5325--5334.
	
	\bibitem{Ahmed2015CVPR}
	E.~Ahmed, M.~Jones, and T.~K. Marks, ``An improved deep learning architecture
	for person re-identification,'' in \emph{IEEE Conference on Computer Vision
		and Pattern Recognition (CVPR)}, 2015, pp. 3908--3916.
	
	\bibitem{Zhou2017CVPR}
	S.~Zhou, J.~Wang, J.~Wang, Y.~Gong, and N.~Zheng, ``Point to set similarity
	based deep feature learning for person re-identification,'' in \emph{IEEE
		Conference on Computer Vision and Pattern Recognition (CVPR)}, 2017, pp.
	3741--3750.
	
	\bibitem{Ronneberger2015MICCAI}
	O.~Ronneberger, P.~Fischer, and T.~Brox, ``{U-Net}: Convolutional networks for
	biomedical image segmentation,'' in \emph{International Conference on Medical
		Image Computing and Computer-Assisted Intervention}.\hskip 1em plus 0.5em
	minus 0.4em\relax Springer, 2015, pp. 234--241.
	
	\bibitem{Kiranyaz2015TBE}
	S.~Kiranyaz, T.~Ince, and M.~Gabbouj, ``Real-time patient-specific ecg
	classification by 1-d convolutional neural networks,'' \emph{IEEE
		Transactions on Biomedical Engineering}, vol.~63, no.~3, pp. 664--675, 2015.
	
	\bibitem{Simonyan2014arXiv}
	K.~Simonyan and A.~Zisserman, ``Very deep convolutional networks for
	large-scale image recognition,'' \emph{arXiv Preprint: 1409.1556}, 2014.
	
	\bibitem{Mikolov2013arXiv}
	T.~Mikolov, K.~Chen, G.~Corrado, and J.~Dean, ``Efficient estimation of word
	representations in vector space,'' \emph{arXiv preprint: 1301.3781}, 2013.
	
	\bibitem{Yan2017NAT}
	G.~Yan, P.~E. V{\'e}rtes, E.~K. Towlson, Y.~L. Chew, D.~S. Walker, W.~R.
	Schafer, and A.-L. Barab{\'a}si, ``Network control principles predict neuron
	function in the caenorhabditis elegans connectome,'' \emph{Nature}, vol. 550,
	no. 7677, p. 519, 2017.
	
	\bibitem{Glorot2011ICAIS}
	X.~Glorot, A.~Bordes, and Y.~Bengio, ``Deep sparse rectifier neural networks,''
	in \emph{International Conference on Artificial Intelligence and Statistics},
	2011, pp. 315--323.
	
	\bibitem{Erdos1964RG}
	P.~Erd{\"{o}}s and A.~R{\'e}nyi, ``On the strength of connectedness of a random
	graph,'' \emph{Acta Mathematica Hungarica}, vol.~12, no. 1-2, pp. 261--267,
	1964.
	
	\bibitem{Goh2001PRL}
	K.-I. Goh, B.~Kahng, and D.~Kim, ``Universal behavior of load distribution in
	scale-free networks,'' \emph{Physical Review Letters}, vol.~87, no.~27, p.
	278701, 2001.
	
	\bibitem{Sorrentino2007CH}
	F.~Sorrentino, ``Effects of the network structural properties on its
	controllability,'' \emph{Chaos: An Interdisciplinary Journal of Nonlinear
		Science}, vol.~17, no.~3, p. 033101, 2007.
	
	\bibitem{Newman1999PLA}
	M.~E. Newman and D.~J. Watts, ``Renormalization group analysis of the
	small-world network model,'' \emph{Physics Letters A}, vol. 263, no. 4-6, pp.
	341--346, 1999.
	
	\bibitem{Niepert2016ICML}
	M.~Niepert, M.~Ahmed, and K.~Kutzkov, ``Learning convolutional neural networks
	for graphs,'' in \emph{International Conference on Machine Learning (ICML)},
	2016, pp. 2014--2023.
	
	\bibitem{Kingma2014arXiv}
	D.~P. Kingma and J.~Ba, ``Adam: A method for stochastic optimization,''
	\emph{arXiv Preprint: 1412.6980}, 2014.
	
	\bibitem{Pan2009TKDE}
	S.~J. Pan and Q.~Yang, ``A survey on transfer learning,'' \emph{IEEE
		Transactions on Knowledge and Data Engineering}, vol.~22, no.~10, pp.
	1345--1359, 2009.
	
\end{thebibliography}
\vfill

\end{document}